\documentclass{article}

\usepackage{microtype}
\usepackage{graphicx}
\usepackage{booktabs} 
\usepackage{caption}
\usepackage{subcaption}

\usepackage{hyperref}

\usepackage[accepted]{icml2023}

\usepackage{amsmath}
\usepackage{amssymb}
\usepackage{mathtools}
\usepackage{amsthm}

\usepackage{soul}
\usepackage{makecell}
\usepackage{array}

\usepackage{natbib}

\usepackage[capitalize,noabbrev]{cleveref}

\theoremstyle{plain}

\theoremstyle{definition}

\theoremstyle{remark}

\usepackage[textsize=tiny]{todonotes}

\icmltitlerunning{Detecting Tidal Features using Self-Supervised Representation Learning}

\begin{document}

\twocolumn[
\icmltitle{Detecting Tidal Features using Self-Supervised Representation Learning}



\icmlsetsymbol{equal}{*}

\begin{icmlauthorlist}
\icmlauthor{Alice Desmons}{UNSW}
\icmlauthor{Sarah Brough}{UNSW}
\icmlauthor{Francois Lanusse}{CEA}
\end{icmlauthorlist}

\icmlaffiliation{UNSW}{School of Physics, University of New South Wales, NSW 2052, Australia}
\icmlaffiliation{CEA}{AIM, CEA, CNRS, Universit\'e Paris-Saclay, Universit\'e Paris Diderot, Sorbonne Paris Cit\'e, F-91191 Gif-sur-Yvette, France}

\icmlcorrespondingauthor{Alice Desmons}{a.desmons@unsw.edu.edu.au}

\icmlkeywords{Machine Learning, ICML}

\vskip 0.3in
]


\printAffiliationsAndNotice{}  

\begin{abstract}
Low surface brightness substructures around galaxies, known as tidal features, are a valuable tool in the detection of past or ongoing galaxy mergers. Their properties can answer questions about the progenitor galaxies involved in the interactions. This paper presents promising results from a self-supervised machine learning model, trained on data from the Ultradeep layer of the Hyper Suprime-Cam Subaru Strategic Program optical imaging survey, designed to automate the detection of tidal features. We find that self-supervised models are capable of detecting tidal features and that our model outperforms previous automated tidal feature detection methods, including a fully supervised model. The previous state of the art method achieved 76\% completeness for 22\% contamination, while our model achieves considerably higher (96\%) completeness for the same level of contamination. 
\end{abstract}

\section{Introduction}
\label{sec:intro}

The currently accepted model of the Universe, known as the Lambda Cold Dark Matter ($\Lambda$CDM) Cosmological Model, postulates that galaxies evolve through a process which is referred to as the `hierarchical merger model’, wherein the growth of the universe's highest-mass galaxies is dominated by merging with lower-mass galaxies (e.g. \citealt{Lacey1994NBodyMergeRate, Cole2000Hierarchical, Robotham2014GAMAClosePair, Martin2018MergeMorphTransform}). During the merging process, the extreme gravitational forces involved cause stellar material to be pulled out from the galaxies, forming diffuse non-uniform regions of stars in the outskirts of the galaxies, known as tidal features. These tidal features contain information about the merging history of the galaxy, and can thus be used to study the galaxy evolution process. 

In order to draw accurate and statistically robust conclusions about this evolution process, we require a large sample of galaxies exhibiting tidal features. One thing that makes this difficult is the extremely low surface brightness of tidal features, which can easily reach $\mu_{r}\geq$~27~mag~$\rm{arcsec}^{-2}$. With the next generation of wide-field optical imaging surveys reaching new limiting depths, such as the Vera C Rubin Observatory's Legacy Survey of Space and Time (LSST; \citealt{Ivezic2019LSST}) which is predicted to reach $\mu_{r}\sim$~30.1~mag~$\rm{arcsec}^{-2}$ \citep{Martin2022TidalFeatMockIm}, assembling a statistically significant sample of galaxies with tidal features is becoming more feasible. One challenge associated with surveys like LSST, due to commence in 2024 and run for 10 years, is the amount of data predicted to be released, with LSST predicted to output over 500 petabytes of imaging data including billions of galaxies \citep{Ivezic2019LSST}. Current tidal feature detection and classification is primarily achieved through visual identification (e.g. \citealt{Tal2009EllipGalTidalFeat, Sheen2012PostMergeSigs, Atkinson2013CFHTLSTidal, Hood2018RESOLVETidalFeat, Bilek2020MATLASTidalFeat, Martin2022TidalFeatMockIm}), but this amount of data is virtually impossible to classify visually by humans, even using large community based projects such as Galaxy Zoo \citep{Lintott2008GalZoo, Darg2010GalZooFracMarge}, and hence we are in urgent need of a tool that can automate this classification task and isolate galaxies with tidal features.

With the promising recent results of machine learning in galaxy classification tasks (e.g. \citealt{Hocking2018UnsupGalMorph,Diaz2019CNNGalFormProcess,Pearson2019DeepLearnMergers,Snyder2019IllustrisAutoMergerClass,Walmsley2019CNNTidalFeat,Cavanagh2020DeepLearnBars,Martin2020UnsupMorphClass}), we turn to machine learning to construct a model which can take galaxy images as input, convert them into representations - low-dimensional maps which preserve the important information in the image - and output a classification based on whether the galaxy possesses tidal features. We use a recently developed machine learning method that is essentially a middle-point between supervised and unsupervised learning, known as Self-Supervised machine Learning (SSL; \citealt{He2019UnsupMomentContrast, ChenT2020ContrastiveFrame,ChenT2020SelfSup, ChenX2020MomentContrastive, ChenX2020SimSiam}). Such models do not require labelled data for the training of the encoder, which learns to transform images into meaningful low-dimensional representations, but can perform classification when paired with a linear classifier and a small labelled dataset. Instead of labels, SSL models rely on augmentations to learn under which conditions the output low-dimensional representations should be invariant. These types of models have been successfully used for a variety of astronomical applications (e.g. \citealt{Hayat2021SSMLAstroIms,Stein2022SelfSupGravLens,Slijepcevic2022SSMLRadio,Walmsley2022SSMLGalMorph,Wei2022SSMLMorph,Ciprijanovic2023SSMLCrossSurv,Huertas-Company2023ContLearn,Slijepcevic2023SSMLRadio}) Compared to supervised models, self-supervised models are also much easier to adapt to perform new tasks, and apply to datasets from different astronomical surveys \citep{Ciprijanovic2023SSMLCrossSurv} making this kind of model perfect for our goal of applying the tool developed using HSC-SSP data to future LSST data.

\section{Methods}
\label{sec:Methods}
\subsection{Sample Selection}
\label{sec:data}
The dataset used for this work is sourced from the Ultradeep (UD) layer of the HSC-SSP Public Data Release 2 (PDR2; \citealt{Aihara2019HSCSecondData}) for deep galaxy images. We use the Ultradeep field, which spans an area of $3.5$~deg$^{2}$ and reaches a surface brightness depth of $\mu_{r}\sim$ 28.0~mag arcsec$^{-2}$ as it reaches depths faint enough to detect tidal features.

We assemble an unlabelled dataset of $\sim$44,000 galaxies by parsing objects in the HSC-SSP PDR2 database using an SQL search and only selecting objects which have at least 3 exposures in each band and have \textit{i}-band magnitudes 15~$<$~\textit{i}~$<$~20 mag. We set a faint magnitude limit of 20 mag to ensure that objects are bright enough for tidal features to be visible. We access the HSC-SSP galaxy images using the ‘Unagi’ Python tool \citep{Huang2019Unagi} which, given a galaxy’s right ascension and declination, allows us to create multi-band ‘HSC cutout’ images of size 128~$\times$~128 pixels, or 21~$\times$~21 arcsecs, centred around each galaxy. Each cutout is downloaded in five ($g,~r,~i,~z,~y$) bands.

For the training of the linear classifier we require a small labelled dataset of galaxies with and without tidal features. We use the HSC-SSP UD PDR2 dataset assembled by \citet{Desmons2023GAMA} composed of 211 galaxies with tidal features and 641 galaxies without tidal features. These galaxies were selected from a volume-limited sample from the cross-over between then Galaxy and Mass Assembly survey \citep{Driver2011GAMADataRel} and HSC-SSP with spectroscopic redshift limits 0.04~$\leq$~\textit{z}~$\leq$~0.2 and stellar mass limits 9.50~$\leq$~log$_{10}$($M_{\star}$/M$_{\odot}$)~$\leq$~11.00 and have \textit{i}-band magnitudes in the range 12.8~$<$~\textit{i}~$<$~21.6~mag. To increase the size of our tidal feature training sample we classified additional galaxies from our HSC-SSP PDR2 unlabelled dataset of $\sim$~44,000 objects, according to the classification scheme outlined in \citet{Desmons2023GAMA}. Our final labelled sample contains 760 galaxies, 380 with tidal features, labelled 1, and 380 without, labelled 0. We split our labelled dataset set into training, validation, and testing datasets composed of 600, 60, and 100 galaxies respectively. 

\subsection{Image Pre-processing and Augmentations}
\label{sec:augs}
Before the images are augmented and fed through the model we apply a pre-processing function to normalise the images. The augmentations we use for this project are:
\begin{itemize}
    \setlength\itemsep{0.1em}
    \item \textbf{Orientation:} We randomly flip the image across each axis (x and y) with 50\% probability.
    \item \textbf{Gaussian Noise}: We sample a scalar from $\mathcal{U}$(1,3) and multiply it with the median absolute deviation of each channel (calculated over 1000 training examples) to get a per-channel noise $\sigma_{c}$. We then introduce Gaussian noise sampled from $\sigma_{c}~\times~\mathcal{N}$(0,1) for each channel.
    \item \textbf{Jitter and Crop:} For HSC-SSP images we crop the 128~$\times$~128 pixel image to the central 109~$\times$~109 pixels before randomly cropping the image to 96~$\times$~96 pixel. Random cropping means the image center is translated, or `jittered', along each respective axis by $i$, $j$ pixels where $i$, $j~\sim~\mathcal{U}$(-13,13) before cropping to the central 96~$\times$~96 pixels.
\end{itemize}

\subsection{Model Architecture}
\label{sec:mod_arch}
The model we utilise to perform classification of tidal feature candidates consists of two components; a self-supervised model used for pre-training, and a `fine-tuned' model used for classification. All models described below are built using the TensorFlow framework \citep{Abadi2016TensorFlow}.

\subsubsection{The Self-Supervised Architecture}
\label{sec:self_sup_arch}
For our task of classifying tidal feature candidates we use a type of self-supervised learning known as Nearest Neighbour Contrastive Learning of visual Representations (NNCLR; \citealt{Dwibedi2021NNCLR}). We closely follow \citet{Dwibedi2021NNCLR} in designing the architecture and training process for our model. The model was compiled using the Adam optimiser \citep{Kingma2014AdamLoss} and trained for 25 epochs on our unlabelled dataset of $\sim$~44,000 HSC-SSP PDR2 galaxies. 

\subsubsection{The Fine-tuned Architecture}
\label{sec:fine_tune_arch}
The fine-tuned model is a simple linear classifier which takes galaxy images as input and converts them to representations using the pre-trained self-supervised encoder. These representations are passed through a `Dense' layer with a sigmoid activation, which outputs a single number between 0 and 1. This fine-tuned model was compiled using the Adam optimiser \citep{Kingma2014AdamLoss} and a binary cross entropy loss. It was trained for 50 epochs using the labelled training set of 600 HSC-SSP galaxies. Training was completed within $\sim$~1 minute using a single GPU.

\subsubsection{The Supervised Architecture}
\label{sec:sup_arch}
To draw conclusions about the suitability of self-supervised models for the detection and classification of tidal features, we compare our results with those of a fully supervised model. We do not construct this model from scratch, but instead use the published model designed by \citet{Pearson2019DeepLearnMergers} to classify merging galaxies. The output layer was changed from two neurons with softmax activation, to a single neuron with sigmoid activation. The network was compiled using the Adam optimiser \citep{Kingma2014AdamLoss} with the default learning rate and loss of the network was determined using binary cross entropy. We additionally changed the input image dimension from 64~$\times$~64 pixels with three colour channels to 96~$\times$~96 pixels with five colour channels to ensure extended tidal features remain visible. We train this fully supervised network from scratch using the labelled training set of 600 HSC-SSP galaxies.

\subsection{Model Evaluation}
\label{sec:mod_eval}
To evaluate our model performance we use the true positive rate (also known as recall or completeness) and false positive rate (also known as fall-out or contamination). The true positive rate (TPR) ranges from 0 to 1 and is defined as the fraction of galaxies correctly classified by the model as having tidal features with respect to the total number of galaxies with tidal features. The false positive rate (FPR) also ranges from 0 to 1 and is defined as the fraction of galaxies incorrectly classified by the model as having tidal features with respect to the total number of galaxies without tidal features.

In addition to using the TPR for a given FPR to evaluate our model, we also use the area under the receiver operating characteristic (ROC) curve, or ROC AUC, to evaluate performance.

\section{Results}
\label{sec:results}
\subsection{Self-Supervised vs. Supervised Performance}
\label{sec:res_comp}

\begin{figure}[h]
    \vskip 0.2in
    \begin{center}
    \includegraphics[width=0.8\columnwidth]{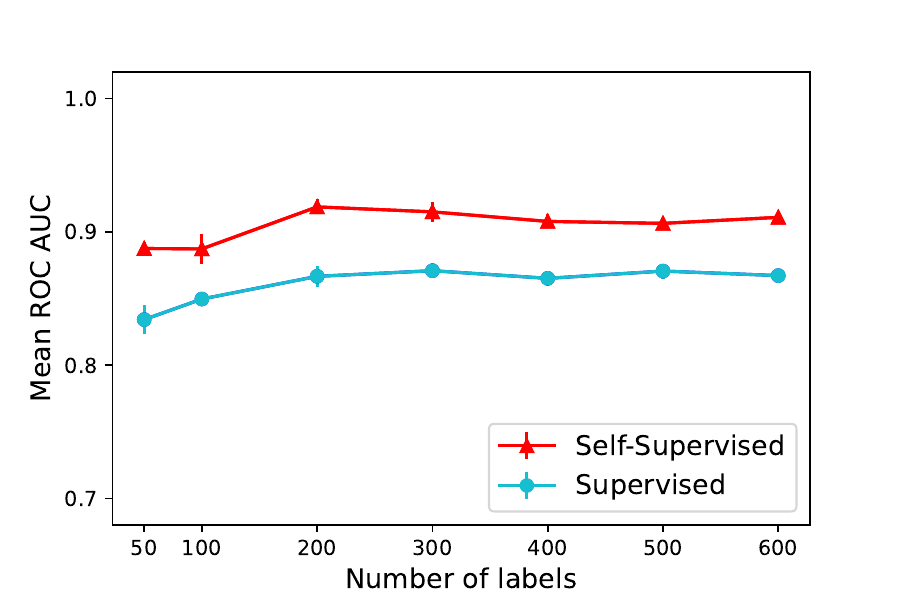}
    \caption{Average ROC AUC as a function of the number of HSC-SSP labels used for training for a supervised (blue) and self-supervised (red) model. Each point is an average of ten runs.}
    \label{fig:HSC_AUC}
    \end{center}
    \vskip -0.2in
\end{figure}

Figure \ref{fig:HSC_AUC} illustrates the testing set ROC AUC for a supervised and self-supervised network as a function of the number of labels used in training for our HSC-SSP dataset. Each point represents the ROC AUC averaged over ten runs using the same training, validation, and testing sets for each run. We average the ROC AUC over the 10 runs and remove outliers further than $3\sigma$ from the mean. Our SSL model maintains high performance across all amounts of labels used for training, having ROC~AUC~$=$~0.911~$\pm$~0.002 when training on the maximum number of labels and only dropping to ROC~AUC~$=$~0.89~$\pm$~0.01 when using only 50 labels for training. The supervised model also maintains its performance regardless of label number, but only reaches ROC~AUC~$=$~0.867~$\pm$~0.004 when training on the maximum number and ROC~AUC~$=$~0.83~$\pm$~0.01 when using 50 labels for training. 
This figure not only shows that an SSL model can be used for the detection of tidal features with good performance, but also that it performs consistently better than the supervised network regardless of the number of training labels. We also calculated the average TPR reached by the self-supervised model on the testing set for a given FPR~$=$~0.2, averaging over 10 runs and removing outliers. When training using 600 labels, the model reaches TPR~$=$~0.94~$\pm$~0.01, and this only drops to TPR~$=$~0.90~$\pm$~0.01 when using a mere 50 labels for training. 

\begin{figure*}[t]
  \vskip 0.2in
  \begin{center}
  \includegraphics[width=0.7\textwidth,]{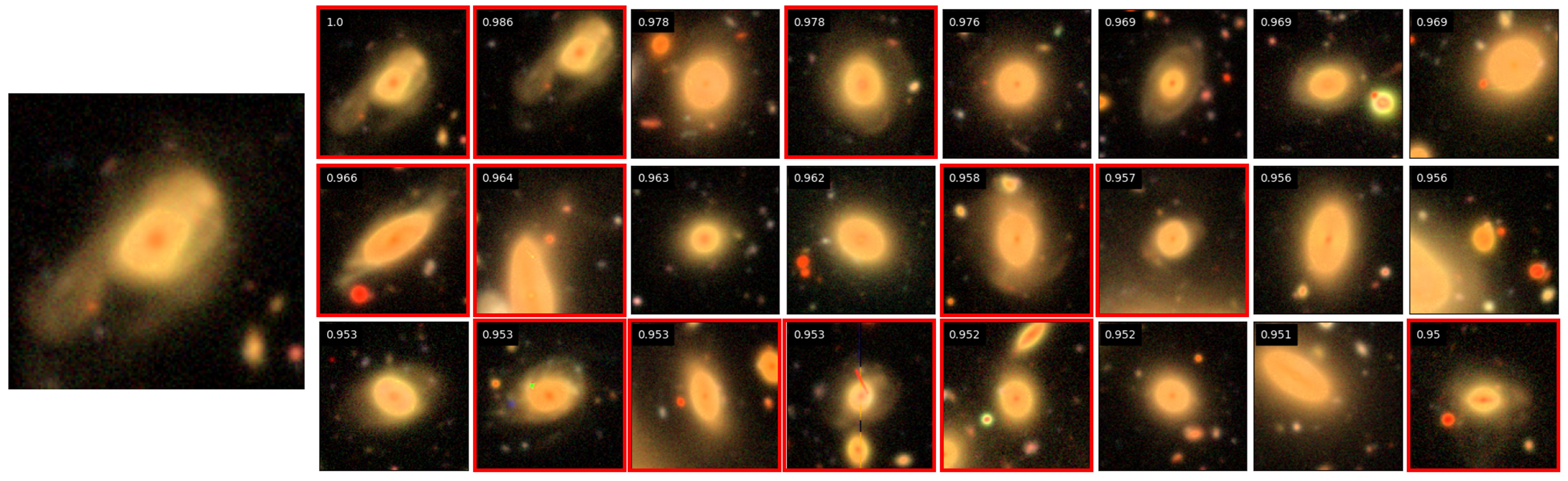}
  \caption{Results from a similarity search using a random galaxy with tidal features as a query image, displayed on the left, alongside the top 24 galaxies with the highest similarity scores for each similarity search on the right. The similarity score is displayed in the top left corner for each image. The red outlines indicate images containing galaxies which would be visually classified as hosting tidal features, regardless of whether this galaxy is the central object in the image.}
  \label{fig:HSC_sim_search}
  \end{center}
  \vskip -0.2in
\end{figure*}
\begin{figure*}
  \vskip 0.2in
  \begin{center}
  \includegraphics[width=0.8\textwidth]{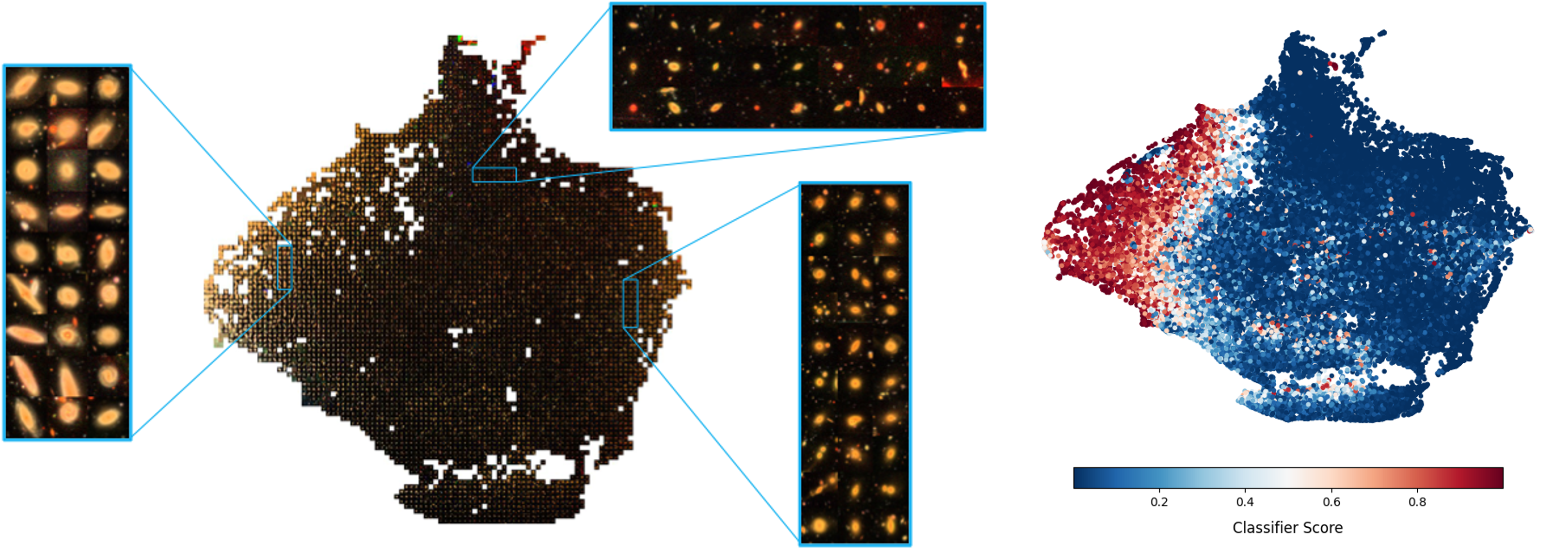}
  \caption{Left: 2D UMAP projection of the self-supervised representations. Made by binning the space into $100~\times~100$ cells and randomly selecting a sample from that cell to plot in the corresponding cell location. Right: The same 2D UMAP projection without binning, coloured according the scores assigned to each galaxy by the linear classifier.}
  \label{fig:UMAP_preview}
  \end{center}
  \vskip -0.2in
\end{figure*}
\subsection{Detection of Tidal Features}
\label{sec:res_HSC}
One advantage of self-supervised models over supervised models is the ability to use just one labelled example to find examples of similar galaxies from the full dataset. By using just one image from our labelled tidal feature dataset as a query image, and the encoded 128-dimensional representations from the self-supervised encoder, we can perform a similarity search that assigns high similarity scores to images which have similar representations to the query image. This is demonstrated in Figure \ref{fig:HSC_sim_search} where we select a random galaxy with tidal features from our training sample and perform a similarity search with the 44,000 unlabelled HSC-SSP galaxies. In Figure \ref{fig:HSC_sim_search} the query image is shown on the right alongside the 24 galaxies which received the highest similarity scores. This figure shows the power of self-supervised learning, where using only a single labelled example, we can find a multitude of other tidal feature candidates.

We can also visualise how the model organises the galaxy images in representation space, by using Uniform Manifold Approximation and Projection (UMAP; \citealt{McInnes2018UMAP}) which reduces the encoded representations to an easier to visualise 2 dimensional projection. Figure \ref{fig:UMAP_preview} illustrates this 2D projection, created by binning the space into $100~\times~100$ cells and randomly selecting a sample from that cell to plot in the corresponding cell location. We also enquire whether the scores given to galaxies by the linear classifier are related to the galaxies' positions in the UMAP projection, by colouring the UMAP plot according the scores given to each galaxy by the linear classifier, shown in the right panel of Figure \ref{fig:UMAP_preview}. We find that the majority of galaxies which were assigned a high classifier score, indicating a high likelihood of tidal features, are located on the left side of the UMAP projection plot. This reinforces the idea that the encoded representations contain meaningful information about tidal features.

\section{Discussion and Conclusions}
\label{sec:disc}

In this work, we have shown that SSL models composed of a self-supervised encoder and linear classifier can not only be used to detect galaxies with tidal features, but can do so reaching both high completeness (TPR~$=$~0. 94~$\pm$~0.1) for low contamination (FPR~$=$~0.20) and high area under the ROC curve (ROC~AUC~$=$~0.91~$\pm$~0.002). This means that such models can be used to isolate the majority of galaxies with tidal features from a large sample of galaxies, thus drastically reducing the amount of visual classification needed to assemble a large sample of tidal features. One major advantage of this model over other automated classification methods, is that this level of performance can be reached using only 600 labelled training examples, and only drops mildly when using a mere 50 labels for training maintaining ROC~AUC~$=$~0.89~$\pm$~0.01 and TPR~$=$~0.90~$\pm$~0.1 for FPR~$=$~0.2. This makes SSL models easy to re-train on data from different surveys with minimal visual classification needed. Following \citet{Stein2021SelfSupSim}, we emphasise the usefulness of being able to perform a similarity search using just the self-supervised encoder and one example of a galaxy with tidal features to find other galaxies with tidal features from a dataset of tens of thousands of galaxies.  

The level of comparison that can be carried out with respect to the results obtained here and other works is limited due to the scarcity of similar works. There is only one study focusing on the detection of tidal features using machine learning, namely the work of \citet{Walmsley2019CNNTidalFeat} who used a supervised network to identify galaxies with tidal features from the Wide layer of the Canada-France-Hawaii Telescope Legacy Survey \citep{Gwyn2012CFHTLS}. \citet{Walmsley2019CNNTidalFeat} found that their method outperformed other automated methods of tidal feature detection, reaching 76\% completeness (or TPR) and 22\% contamination (or FPR). Our SSL model, trained on 600 galaxies performs considerably better, reaching a completeness of 96\% for the same contamination percentage. Most importantly, our model consistently outperforms a fully supervised model trained on the same data, reaching ROC AUC~=~0.911~$\pm$0.002 while the fully supervised model only reaches a maximum ROC AUC of 0.864~$\pm$~0.004. 

The code use to create, train, validate, and test the SSML model, along with instructions on loading and using the pre-trained model as well as training the model using different data can be downloaded from GitHub\footnote{\url{https://github.com/LSSTISSC/Tidalsaurus}}.


\newpage
\bibliography{main}

\begin{thebibliography}{}

\bibitem[{Abadi} et~al., 2016]{Abadi2016TensorFlow}
{Abadi}, M., {Agarwal}, A., {Barham}, P., {Brevdo}, E., {Chen}, Z., {Citro},
  C., {Corrado}, G.~S., {Davis}, A., {Dean}, J., {Devin}, M., {Ghemawat}, S.,
  {Goodfellow}, I., {Harp}, A., {Irving}, G., {Isard}, M., {Jia}, Y.,
  {Jozefowicz}, R., {Kaiser}, L., {Kudlur}, M., {Levenberg}, J., {Mane}, D.,
  {Monga}, R., {Moore}, S., {Murray}, D., {Olah}, C., {Schuster}, M., {Shlens},
  J., {Steiner}, B., {Sutskever}, I., {Talwar}, K., {Tucker}, P., {Vanhoucke},
  V., {Vasudevan}, V., {Viegas}, F., {Vinyals}, O., {Warden}, P., {Wattenberg},
  M., {Wicke}, M., {Yu}, Y., and {Zheng}, X. (2016).
\newblock {TensorFlow: Large-Scale Machine Learning on Heterogeneous
  Distributed Systems}.
\newblock {\em arXiv e-prints}, page arXiv:1603.04467.

\bibitem[{Aihara} et~al., 2019]{Aihara2019HSCSecondData}
{Aihara}, H., {AlSayyad}, Y., {Ando}, M., {Armstrong}, R., {Bosch}, J.,
  {Egami}, E., {Furusawa}, H., {Furusawa}, J., {Goulding}, A., {Harikane}, Y.,
  {Hikage}, C., {Ho}, P. T.~P., {Hsieh}, B.-C., {Huang}, S., {Ikeda}, H.,
  {Imanishi}, M., {Ito}, K., {Iwata}, I., {Jaelani}, A.~T., {Kakuma}, R.,
  {Kawana}, K., {Kikuta}, S., {Kobayashi}, U., {Koike}, M., {Komiyama}, Y.,
  {Li}, X., {Liang}, Y., {Lin}, Y.-T., {Luo}, W., {Lupton}, R., {Lust}, N.~B.,
  {MacArthur}, L.~A., {Matsuoka}, Y., {Mineo}, S., {Miyatake}, H., {Miyazaki},
  S., {More}, S., {Murata}, R., {Namiki}, S.~V., {Nishizawa}, A.~J., {Oguri},
  M., {Okabe}, N., {Okamoto}, S., {Okura}, Y., {Ono}, Y., {Onodera}, M.,
  {Onoue}, M., {Osato}, K., {Ouchi}, M., {Shibuya}, T., {Strauss}, M.~A.,
  {Sugiyama}, N., {Suto}, Y., {Takada}, M., {Takagi}, Y., {Takata}, T.,
  {Takita}, S., {Tanaka}, M., {Terai}, T., {Toba}, Y., {Uchiyama}, H.,
  {Utsumi}, Y., {Wang}, S.-Y., {Wang}, W., and {Yamada}, Y. (2019).
\newblock {Second data release of the Hyper Suprime-Cam Subaru Strategic
  Program}.
\newblock {\em PASJ}, 71(6):114.

\bibitem[{Atkinson} et~al., 2013]{Atkinson2013CFHTLSTidal}
{Atkinson}, A.~M., {Abraham}, R.~G., and {Ferguson}, A. M.~N. (2013).
\newblock {Faint Tidal Features in Galaxies within the Canada-France-Hawaii
  Telescope Legacy Survey Wide Fields}.
\newblock {\em ApJ}, 765(1):28.

\bibitem[{B{\'\i}lek} et~al., 2020]{Bilek2020MATLASTidalFeat}
{B{\'\i}lek}, M., {Duc}, P.-A., {Cuillandre}, J.-C., {Gwyn}, S., {Cappellari},
  M., {Bekaert}, D.~V., {Bonfini}, P., {Bitsakis}, T., {Paudel}, S.,
  {Krajnovi{\'c}}, D., {Durrell}, P.~R., and {Marleau}, F. (2020).
\newblock {Census and classification of low-surface-brightness structures in
  nearby early-type galaxies from the MATLAS survey}.
\newblock {\em MNRAS}, 498(2):2138--2166.

\bibitem[{Cavanagh} and {Bekki}, 2020]{Cavanagh2020DeepLearnBars}
{Cavanagh}, M.~K. and {Bekki}, K. (2020).
\newblock {Bars formed in galaxy merging and their classification with deep
  learning}.
\newblock {\em A\&A}, 641:A77.

\bibitem[{Chen} et~al., 2020a]{ChenT2020ContrastiveFrame}
{Chen}, T., {Kornblith}, S., {Norouzi}, M., and {Hinton}, G. (2020a).
\newblock {A Simple Framework for Contrastive Learning of Visual
  Representations}.
\newblock {\em arXiv e-prints}, page arXiv:2002.05709.

\bibitem[{Chen} et~al., 2020b]{ChenT2020SelfSup}
{Chen}, T., {Kornblith}, S., {Swersky}, K., {Norouzi}, M., and {Hinton}, G.
  (2020b).
\newblock {Big Self-Supervised Models are Strong Semi-Supervised Learners}.
\newblock {\em arXiv e-prints}, page arXiv:2006.10029.

\bibitem[{Chen} et~al., 2020c]{ChenX2020MomentContrastive}
{Chen}, X., {Fan}, H., {Girshick}, R., and {He}, K. (2020c).
\newblock {Improved Baselines with Momentum Contrastive Learning}.
\newblock {\em arXiv e-prints}, page arXiv:2003.04297.

\bibitem[{Chen} and {He}, 2020]{ChenX2020SimSiam}
{Chen}, X. and {He}, K. (2020).
\newblock {Exploring Simple Siamese Representation Learning}.
\newblock {\em arXiv e-prints}, page arXiv:2011.10566.

\bibitem[{{\'C}iprijanovi{\'c}} et~al., 2023]{Ciprijanovic2023SSMLCrossSurv}
{{\'C}iprijanovi{\'c}}, A., {Lewis}, A., {Pedro}, K., {Madireddy}, S., {Nord},
  B., {Perdue}, G.~N., and {Wild}, S.~M. (2023).
\newblock {DeepAstroUDA: Semi-Supervised Universal Domain Adaptation for
  Cross-Survey Galaxy Morphology Classification and Anomaly Detection}.
\newblock {\em arXiv e-prints}, page arXiv:2302.02005.

\bibitem[{Cole} et~al., 2000]{Cole2000Hierarchical}
{Cole}, S., {Lacey}, C.~G., {Baugh}, C.~M., and {Frenk}, C.~S. (2000).
\newblock {Hierarchical galaxy formation}.
\newblock {\em MNRAS}, 319(1):168--204.

\bibitem[{Darg} et~al., 2010]{Darg2010GalZooFracMarge}
{Darg}, D.~W., {Kaviraj}, S., {Lintott}, C.~J., {Schawinski}, K., {Sarzi}, M.,
  {Bamford}, S., {Silk}, J., {Proctor}, R., {Andreescu}, D., {Murray}, P.,
  {Nichol}, R.~C., {Raddick}, M.~J., {Slosar}, A., {Szalay}, A.~S., {Thomas},
  D., and {Vandenberg}, J. (2010).
\newblock {Galaxy Zoo: the fraction of merging galaxies in the SDSS and their
  morphologies}.
\newblock {\em MNRAS}, 401(2):1043--1056.

\bibitem[{Desmons} et~al., 2023]{Desmons2023GAMA}
{Desmons}, A., {Brough}, S., {Mart{\'\i}nez-Lombilla}, C., {De Propris}, R.,
  {Holwerda}, B., and {L{\'o}pez S{\'a}nchez}, {\'A}.~R. (2023).
\newblock {Galaxy and mass assembly (GAMA): Comparing visually and
  spectroscopically identified galaxy merger samples}.
\newblock {\em MNRAS}.

\bibitem[{Diaz} et~al., 2019]{Diaz2019CNNGalFormProcess}
{Diaz}, J.~D., {Bekki}, K., {Forbes}, D.~A., {Couch}, W.~J., {Drinkwater},
  M.~J., and {Deeley}, S. (2019).
\newblock {Classifying the formation processes of S0 galaxies using
  Convolutional Neural Networks}.
\newblock {\em MNRAS}, 486(4):4845--4862.

\bibitem[{Driver} et~al., 2011]{Driver2011GAMADataRel}
{Driver}, S.~P., {Hill}, D.~T., {Kelvin}, L.~S., {Robotham}, A.~S.~G., {Liske},
  J., {Norberg}, P., {Baldry}, I.~K., {Bamford}, S.~P., {Hopkins}, A.~M.,
  {Loveday}, J., {Peacock}, J.~A., {Andrae}, E., {Bland-Hawthorn}, J.,
  {Brough}, S., {Brown}, M.~J.~I., {Cameron}, E., {Ching}, J.~H.~Y., {Colless},
  M., {Conselice}, C.~J., {Croom}, S.~M., {Cross}, N.~J.~G., {de Propris}, R.,
  {Dye}, S., {Drinkwater}, M.~J., {Ellis}, S., {Graham}, A.~W., {Grootes},
  M.~W., {Gunawardhana}, M., {Jones}, D.~H., {van Kampen}, E., {Maraston}, C.,
  {Nichol}, R.~C., {Parkinson}, H.~R., {Phillipps}, S., {Pimbblet}, K.,
  {Popescu}, C.~C., {Prescott}, M., {Roseboom}, I.~G., {Sadler}, E.~M.,
  {Sansom}, A.~E., {Sharp}, R.~G., {Smith}, D.~J.~B., {Taylor}, E., {Thomas},
  D., {Tuffs}, R.~J., {Wijesinghe}, D., {Dunne}, L., {Frenk}, C.~S., {Jarvis},
  M.~J., {Madore}, B.~F., {Meyer}, M.~J., {Seibert}, M., {Staveley-Smith}, L.,
  {Sutherland}, W.~J., and {Warren}, S.~J. (2011).
\newblock {Galaxy and Mass Assembly (GAMA): survey diagnostics and core data
  release}.
\newblock {\em MNRAS}, 413(2):971--995.

\bibitem[{Dwibedi} et~al., 2021]{Dwibedi2021NNCLR}
{Dwibedi}, D., {Aytar}, Y., {Tompson}, J., {Sermanet}, P., and {Zisserman}, A.
  (2021).
\newblock {With a Little Help from My Friends: Nearest-Neighbor Contrastive
  Learning of Visual Representations}.
\newblock {\em arXiv e-prints}, page arXiv:2104.14548.

\bibitem[Gwyn, 2012]{Gwyn2012CFHTLS}
Gwyn, S. D.~J. (2012).
\newblock The canada-france-hawaii telescope legacy survey: Stacked images and
  catalogs.
\newblock {\em AJ}, 143(2).

\bibitem[{Hayat} et~al., 2021]{Hayat2021SSMLAstroIms}
{Hayat}, M.~A., {Stein}, G., {Harrington}, P., {Luki{\'c}}, Z., and {Mustafa},
  M. (2021).
\newblock {Self-supervised Representation Learning for Astronomical Images}.
\newblock {\em ApJl}, 911(2):L33.

\bibitem[{He} et~al., 2019]{He2019UnsupMomentContrast}
{He}, K., {Fan}, H., {Wu}, Y., {Xie}, S., and {Girshick}, R. (2019).
\newblock {Momentum Contrast for Unsupervised Visual Representation Learning}.
\newblock {\em arXiv e-prints}, page arXiv:1911.05722.

\bibitem[{Hocking} et~al., 2018]{Hocking2018UnsupGalMorph}
{Hocking}, A., {Geach}, J.~E., {Sun}, Y., and {Davey}, N. (2018).
\newblock {An automatic taxonomy of galaxy morphology using unsupervised
  machine learning}.
\newblock {\em MNRAS}, 473(1):1108--1129.

\bibitem[{Hood} et~al., 2018]{Hood2018RESOLVETidalFeat}
{Hood}, C.~E., {Kannappan}, S.~J., {Stark}, D.~V., {Dell'Antonio}, I.~P.,
  {Moffett}, A.~J., {Eckert}, K.~D., {Norris}, M.~A., and {Hendel}, D. (2018).
\newblock {The Origin of Faint Tidal Features around Galaxies in the RESOLVE
  Survey}.
\newblock {\em ApJ}, 857(2):144.

\bibitem[Huang et~al., 2019]{Huang2019Unagi}
Huang, S., Li, J.-X., Lanusse, F., and Bradshaw, C. (2019).
\newblock {Unagi}.
\newblock \url{https://github.com/dr-guangtou/unagi.git}.

\bibitem[{Huertas-Company} et~al., 2023]{Huertas-Company2023ContLearn}
{Huertas-Company}, M., {Sarmiento}, R., and {Knapen}, J. (2023).
\newblock {A brief review of contrastive learning applied to astrophysics}.
\newblock {\em arXiv e-prints}, page arXiv:2306.05528.

\bibitem[{Ivezi{\'c}} et~al., 2019]{Ivezic2019LSST}
{Ivezi{\'c}}, {\v{Z}}., {Kahn}, S.~M., {Tyson}, J.~A., {Abel}, B., {Acosta},
  E., {Allsman}, R., {Alonso}, D., {AlSayyad}, Y., {Anderson}, S.~F., {Andrew},
  J., {Angel}, J. R.~P., {Angeli}, G.~Z., {Ansari}, R., {Antilogus}, P.,
  {Araujo}, C., {Armstrong}, R., {Arndt}, K.~T., {Astier}, P., {Aubourg},
  {\'E}., {Auza}, N., {Axelrod}, T.~S., {Bard}, D.~J., {Barr}, J.~D., {Barrau},
  A., {Bartlett}, J.~G., {Bauer}, A.~E., {Bauman}, B.~J., {Baumont}, S.,
  {Bechtol}, E., {Bechtol}, K., {Becker}, A.~C., {Becla}, J., {Beldica}, C.,
  {Bellavia}, S., {Bianco}, F.~B., {Biswas}, R., {Blanc}, G., {Blazek}, J.,
  {Blandford}, R.~D., {Bloom}, J.~S., {Bogart}, J., {Bond}, T.~W., {Booth},
  M.~T., {Borgland}, A.~W., {Borne}, K., {Bosch}, J.~F., {Boutigny}, D.,
  {Brackett}, C.~A., {Bradshaw}, A., {Brandt}, W.~N., {Brown}, M.~E.,
  {Bullock}, J.~S., {Burchat}, P., {Burke}, D.~L., {Cagnoli}, G., {Calabrese},
  D., {Callahan}, S., {Callen}, A.~L., {Carlin}, J.~L., {Carlson}, E.~L.,
  {Chandrasekharan}, S., {Charles-Emerson}, G., {Chesley}, S., {Cheu}, E.~C.,
  {Chiang}, H.-F., {Chiang}, J., {Chirino}, C., {Chow}, D., {Ciardi}, D.~R.,
  {Claver}, C.~F., {Cohen-Tanugi}, J., {Cockrum}, J.~J., {Coles}, R.,
  {Connolly}, A.~J., {Cook}, K.~H., {Cooray}, A., {Covey}, K.~R., {Cribbs}, C.,
  {Cui}, W., {Cutri}, R., {Daly}, P.~N., {Daniel}, S.~F., {Daruich}, F.,
  {Daubard}, G., {Daues}, G., {Dawson}, W., {Delgado}, F., {Dellapenna}, A.,
  {de Peyster}, R., {de Val-Borro}, M., {Digel}, S.~W., {Doherty}, P.,
  {Dubois}, R., {Dubois-Felsmann}, G.~P., {Durech}, J., {Economou}, F.,
  {Eifler}, T., {Eracleous}, M., {Emmons}, B.~L., {Fausti Neto}, A.,
  {Ferguson}, H., {Figueroa}, E., {Fisher-Levine}, M., {Focke}, W., {Foss},
  M.~D., {Frank}, J., {Freemon}, M.~D., {Gangler}, E., {Gawiser}, E., {Geary},
  J.~C., {Gee}, P., {Geha}, M., {Gessner}, C. J.~B., {Gibson}, R.~R.,
  {Gilmore}, D.~K., {Glanzman}, T., {Glick}, W., {Goldina}, T., {Goldstein},
  D.~A., {Goodenow}, I., {Graham}, M.~L., {Gressler}, W.~J., {Gris}, P., {Guy},
  L.~P., {Guyonnet}, A., {Haller}, G., {Harris}, R., {Hascall}, P.~A., {Haupt},
  J., {Hernandez}, F., {Herrmann}, S., {Hileman}, E., {Hoblitt}, J., {Hodgson},
  J.~A., {Hogan}, C., {Howard}, J.~D., {Huang}, D., {Huffer}, M.~E.,
  {Ingraham}, P., {Innes}, W.~R., {Jacoby}, S.~H., {Jain}, B., {Jammes}, F.,
  {Jee}, M.~J., {Jenness}, T., {Jernigan}, G., {Jevremovi{\'c}}, D., {Johns},
  K., {Johnson}, A.~S., {Johnson}, M. W.~G., {Jones}, R.~L., {Juramy-Gilles},
  C., {Juri{\'c}}, M., {Kalirai}, J.~S., {Kallivayalil}, N.~J., {Kalmbach}, B.,
  {Kantor}, J.~P., {Karst}, P., {Kasliwal}, M.~M., {Kelly}, H., {Kessler}, R.,
  {Kinnison}, V., {Kirkby}, D., {Knox}, L., {Kotov}, I.~V., {Krabbendam},
  V.~L., {Krughoff}, K.~S., {Kub{\'a}nek}, P., {Kuczewski}, J., {Kulkarni}, S.,
  {Ku}, J., {Kurita}, N.~R., {Lage}, C.~S., {Lambert}, R., {Lange}, T.,
  {Langton}, J.~B., {Le Guillou}, L., {Levine}, D., {Liang}, M., {Lim}, K.-T.,
  {Lintott}, C.~J., {Long}, K.~E., {Lopez}, M., {Lotz}, P.~J., {Lupton}, R.~H.,
  {Lust}, N.~B., {MacArthur}, L.~A., {Mahabal}, A., {Mandelbaum}, R.,
  {Markiewicz}, T.~W., {Marsh}, D.~S., {Marshall}, P.~J., {Marshall}, S.,
  {May}, M., {McKercher}, R., {McQueen}, M., {Meyers}, J., {Migliore}, M.,
  {Miller}, M., {Mills}, D.~J., {Miraval}, C., {Moeyens}, J., {Moolekamp},
  F.~E., {Monet}, D.~G., {Moniez}, M., {Monkewitz}, S., {Montgomery}, C.,
  {Morrison}, C.~B., {Mueller}, F., {Muller}, G.~P., {Mu{\~n}oz Arancibia}, F.,
  {Neill}, D.~R., {Newbry}, S.~P., {Nief}, J.-Y., {Nomerotski}, A., {Nordby},
  M., {O'Connor}, P., {Oliver}, J., {Olivier}, S.~S., {Olsen}, K., {O'Mullane},
  W., {Ortiz}, S., {Osier}, S., {Owen}, R.~E., {Pain}, R., {Palecek}, P.~E.,
  {Parejko}, J.~K., {Parsons}, J.~B., {Pease}, N.~M., {Peterson}, J.~M.,
  {Peterson}, J.~R., {Petravick}, D.~L., {Libby Petrick}, M.~E., {Petry},
  C.~E., {Pierfederici}, F., {Pietrowicz}, S., {Pike}, R., {Pinto}, P.~A.,
  {Plante}, R., {Plate}, S., {Plutchak}, J.~P., {Price}, P.~A., {Prouza}, M.,
  {Radeka}, V., {Rajagopal}, J., {Rasmussen}, A.~P., {Regnault}, N., {Reil},
  K.~A., {Reiss}, D.~J., {Reuter}, M.~A., {Ridgway}, S.~T., {Riot}, V.~J.,
  {Ritz}, S., {Robinson}, S., {Roby}, W., {Roodman}, A., {Rosing}, W.,
  {Roucelle}, C., {Rumore}, M.~R., {Russo}, S., {Saha}, A., {Sassolas}, B.,
  {Schalk}, T.~L., {Schellart}, P., {Schindler}, R.~H., {Schmidt}, S.,
  {Schneider}, D.~P., {Schneider}, M.~D., {Schoening}, W., {Schumacher}, G.,
  {Schwamb}, M.~E., {Sebag}, J., {Selvy}, B., {Sembroski}, G.~H., {Seppala},
  L.~G., {Serio}, A., {Serrano}, E., {Shaw}, R.~A., {Shipsey}, I., {Sick}, J.,
  {Silvestri}, N., {Slater}, C.~T., {Smith}, J.~A., {Smith}, R.~C., {Sobhani},
  S., {Soldahl}, C., {Storrie-Lombardi}, L., {Stover}, E., {Strauss}, M.~A.,
  {Street}, R.~A., {Stubbs}, C.~W., {Sullivan}, I.~S., {Sweeney}, D.,
  {Swinbank}, J.~D., {Szalay}, A., {Takacs}, P., {Tether}, S.~A., {Thaler},
  J.~J., {Thayer}, J.~G., {Thomas}, S., {Thornton}, A.~J., {Thukral}, V.,
  {Tice}, J., {Trilling}, D.~E., {Turri}, M., {Van Berg}, R., {Vanden Berk},
  D., {Vetter}, K., {Virieux}, F., {Vucina}, T., {Wahl}, W., {Walkowicz}, L.,
  {Walsh}, B., {Walter}, C.~W., {Wang}, D.~L., {Wang}, S.-Y., {Warner}, M.,
  {Wiecha}, O., {Willman}, B., {Winters}, S.~E., {Wittman}, D., {Wolff}, S.~C.,
  {Wood-Vasey}, W.~M., {Wu}, X., {Xin}, B., {Yoachim}, P., and {Zhan}, H.
  (2019).
\newblock {LSST: From Science Drivers to Reference Design and Anticipated Data
  Products}.
\newblock {\em ApJ}, 873(2):111.

\bibitem[{Kingma} and {Ba}, 2015]{Kingma2014AdamLoss}
{Kingma}, D.~P. and {Ba}, J. (2015).
\newblock {Adam: A Method for Stochastic Optimization}.
\newblock {\em 3rd International Conference for Learning Representations}, page
  arXiv:1412.6980.

\bibitem[{Lacey} and {Cole}, 1994]{Lacey1994NBodyMergeRate}
{Lacey}, C. and {Cole}, S. (1994).
\newblock {Merger Rates in Hierarchical Models of Galaxy Formation - Part Two -
  Comparison with N-Body Simulations}.
\newblock {\em MNRAS}, 271:676.

\bibitem[{Lintott} et~al., 2008]{Lintott2008GalZoo}
{Lintott}, C.~J., {Schawinski}, K., {Slosar}, A., {Land}, K., {Bamford}, S.,
  {Thomas}, D., {Raddick}, M.~J., {Nichol}, R.~C., {Szalay}, A., {Andreescu},
  D., {Murray}, P., and {Vandenberg}, J. (2008).
\newblock {Galaxy Zoo: morphologies derived from visual inspection of galaxies
  from the Sloan Digital Sky Survey}.
\newblock {\em MNRAS}, 389(3):1179--1189.

\bibitem[{Martin} et~al., 2022]{Martin2022TidalFeatMockIm}
{Martin}, G., {Bazkiaei}, A.~E., {Spavone}, M., {Iodice}, E., {Mihos}, J.~C.,
  {Montes}, M., {Benavides}, J.~A., {Brough}, S., {Carlin}, J.~L., {Collins},
  C.~A., {Duc}, P.~A., {G{\'o}mez}, F.~A., {Galaz}, G., {Hern{\'a}ndez-Toledo},
  H.~M., {Jackson}, R.~A., {Kaviraj}, S., {Knapen}, J.~H.,
  {Mart{\'\i}nez-Lombilla}, C., {McGee}, S., {O'Ryan}, D., {Prole}, D.~J.,
  {Rich}, R.~M., {Rom{\'a}n}, J., {Shah}, E.~A., {Starkenburg}, T.~K.,
  {Watkins}, A.~E., {Zaritsky}, D., {Pichon}, C., {Armus}, L., {Bianconi}, M.,
  {Buitrago}, F., {Bus{\'a}}, I., {Davis}, F., {Demarco}, R., {Desmons}, A.,
  {Garc{\'\i}a}, P., {Graham}, A.~W., {Holwerda}, B., {Hon}, D.~S.~H.,
  {Khalid}, A., {Klehammer}, J., {Klutse}, D.~Y., {Lazar}, I., {Nair}, P.,
  {Noakes-Kettel}, E.~A., {Rutkowski}, M., {Saha}, K., {Sahu}, N., {Sola}, E.,
  {V{\'a}zquez-Mata}, J.~A., {Vera-Casanova}, A., and {Yoon}, I. (2022).
\newblock {Preparing for low surface brightness science with the Vera C. Rubin
  Observatory: Characterization of tidal features from mock images}.
\newblock {\em MNRAS}, 513(1):1459--1487.

\bibitem[{Martin} et~al., 2018]{Martin2018MergeMorphTransform}
{Martin}, G., {Kaviraj}, S., {Devriendt}, J.~E.~G., {Dubois}, Y., and {Pichon},
  C. (2018).
\newblock {The role of mergers in driving morphological transformation over
  cosmic time}.
\newblock {\em MNRAS}, 480(2):2266--2283.

\bibitem[{Martin} et~al., 2020]{Martin2020UnsupMorphClass}
{Martin}, G., {Kaviraj}, S., {Hocking}, A., {Read}, S.~C., and {Geach}, J.~E.
  (2020).
\newblock {Galaxy morphological classification in deep-wide surveys via
  unsupervised machine learning}.
\newblock {\em MNRAS}, 491(1):1408--1426.

\bibitem[{McInnes} et~al., 2018]{McInnes2018UMAP}
{McInnes}, L., {Healy}, J., and {Melville}, J. (2018).
\newblock {UMAP: Uniform Manifold Approximation and Projection for Dimension
  Reduction}.
\newblock {\em arXiv e-prints}, page arXiv:1802.03426.

\bibitem[{Pearson} et~al., 2019]{Pearson2019DeepLearnMergers}
{Pearson}, W.~J., {Wang}, L., {Trayford}, J.~W., {Petrillo}, C.~E., and {van
  der Tak}, F.~F.~S. (2019).
\newblock {Identifying galaxy mergers in observations and simulations with deep
  learning}.
\newblock {\em A\&A}, 626:A49.

\bibitem[{Robotham} et~al., 2014]{Robotham2014GAMAClosePair}
{Robotham}, A.~S.~G., {Driver}, S.~P., {Davies}, L.~J.~M., {Hopkins}, A.~M.,
  {Baldry}, I.~K., {Agius}, N.~K., {Bauer}, A.~E., {Bland-Hawthorn}, J.,
  {Brough}, S., {Brown}, M.~J.~I., {Cluver}, M., {De Propris}, R.,
  {Drinkwater}, M.~J., {Holwerda}, B.~W., {Kelvin}, L.~S., {Lara-Lopez}, M.~A.,
  {Liske}, J., {L{\'o}pez-S{\'a}nchez}, {\'A}.~R., {Loveday}, J., {Mahajan},
  S., {McNaught-Roberts}, T., {Moffett}, A., {Norberg}, P., {Obreschkow}, D.,
  {Owers}, M.~S., {Penny}, S.~J., {Pimbblet}, K., {Prescott}, M., {Taylor},
  E.~N., {van Kampen}, E., and {Wilkins}, S.~M. (2014).
\newblock {Galaxy And Mass Assembly (GAMA): galaxy close pairs, mergers and the
  future fate of stellar mass}.
\newblock {\em MNRAS}, 444(4):3986--4008.

\bibitem[{Sheen} et~al., 2012]{Sheen2012PostMergeSigs}
{Sheen}, Y.-K., {Yi}, S.~K., {Ree}, C.~H., and {Lee}, J. (2012).
\newblock {Post-merger Signatures of Red-sequence Galaxies in Rich Abell
  Clusters at z <\raisebox{-0.5ex}\textasciitilde 0.1}.
\newblock {\em ApJs}, 202(1):8.

\bibitem[{Slijepcevic} et~al., 2022]{Slijepcevic2022SSMLRadio}
{Slijepcevic}, I.~V., {Scaife}, A., {Walmsley}, M., and {Bowles}, M.~R. (2022).
\newblock {Learning useful representations for radio astronomy ``in the wild''
  with contrastive learning}.
\newblock In {\em Machine Learning for Astrophysics}, page~53.

\bibitem[{Slijepcevic} et~al., 2023]{Slijepcevic2023SSMLRadio}
{Slijepcevic}, I.~V., {Scaife}, A. M.~M., {Walmsley}, M., {Bowles}, M., {Wong},
  O.~I., {Shabala}, S.~S., and {White}, S.~V. (2023).
\newblock {Radio Galaxy Zoo: Building a multi-purpose foundation model for
  radio astronomy with self-supervised learning}.
\newblock {\em arXiv e-prints}, page arXiv:2305.16127.

\bibitem[{Snyder} et~al., 2019]{Snyder2019IllustrisAutoMergerClass}
{Snyder}, G.~F., {Rodriguez-Gomez}, V., {Lotz}, J.~M., {Torrey}, P., {Quirk},
  A. C.~N., {Hernquist}, L., {Vogelsberger}, M., and {Freeman}, P.~E. (2019).
\newblock {Automated distant galaxy merger classifications from Space Telescope
  images using the Illustris simulation}.
\newblock {\em MNRAS}, 486(3):3702--3720.

\bibitem[{Stein} et~al., 2022]{Stein2022SelfSupGravLens}
{Stein}, G., {Blaum}, J., {Harrington}, P., {Medan}, T., and {Luki{\'c}}, Z.
  (2022).
\newblock {Mining for Strong Gravitational Lenses with Self-supervised
  Learning}.
\newblock {\em ApJ}, 932(2):107.

\bibitem[{Stein} et~al., 2021]{Stein2021SelfSupSim}
{Stein}, G., {Harrington}, P., {Blaum}, J., {Medan}, T., and {Lukic}, Z.
  (2021).
\newblock {Self-supervised similarity search for large scientific datasets}.
\newblock {\em arXiv e-prints}, page arXiv:2110.13151.

\bibitem[{Tal} et~al., 2009]{Tal2009EllipGalTidalFeat}
{Tal}, T., {van Dokkum}, P.~G., {Nelan}, J., and {Bezanson}, R. (2009).
\newblock {The Frequency of Tidal Features Associated with Nearby Luminous
  Elliptical Galaxies From a Statistically Complete Sample}.
\newblock {\em AJ}, 138(5):1417--1427.

\bibitem[{Walmsley} et~al., 2019]{Walmsley2019CNNTidalFeat}
{Walmsley}, M., {Ferguson}, A. M.~N., {Mann}, R.~G., and {Lintott}, C.~J.
  (2019).
\newblock {Identification of low surface brightness tidal features in galaxies
  using convolutional neural networks}.
\newblock {\em MNRAS}, 483(3):2968--2982.

\bibitem[{Walmsley} et~al., 2022]{Walmsley2022SSMLGalMorph}
{Walmsley}, M., {Slijepcevic}, I., {Bowles}, M.~R., and {Scaife}, A. (2022).
\newblock {Toward Galaxy Foundation Models with Hybrid Contrastive Learning}.
\newblock In {\em Machine Learning for Astrophysics}, page~29.

\bibitem[{Wei} et~al., 2022]{Wei2022SSMLMorph}
{Wei}, S., {Li}, Y., {Lu}, W., {Li}, N., {Liang}, B., {Dai}, W., and {Zhang},
  Z. (2022).
\newblock {Unsupervised Galaxy Morphological Visual Representation with Deep
  Contrastive Learning}.
\newblock {\em PASP}, 134(1041):114508.

\end{thebibliography}
\bibliographystyle{apalike}



\end{document}